# Broadband Terahertz Modulation in Electrostatically-doped Artificial Trilayer Graphene

Ioannis Chatzakis,[1,*] Zhen Li,[1] Alex Benderskii,[2] and Stephen B. Cronin[1,2,3,*]

Departments of [1]Electrical Engineering, [2]Chemistry and [3]Physics,

University of Southern California, Los Angeles, California, CA, 90089, United States

**Abstract:** We report a terahertz optical modulator consisting of randomly stacked trilayer graphene (TLG) deposited on an oxidized silicon substrate by means of THz-Time Domain Spectroscopy (THz-TDS). Here, the gate tuning of the Fermi level of the TLG provides the fundamental basis for the modulation of THz transmission. We measured a 15% change in the THz transmission of this device over a broad frequency range (0.6-1.6THz). We also observed a strong absorption >80% in the time-domain signals and a frequency independence of the conductivity. Furthermore, unlike previous studies, we find that the underlying silicon substrate, which serves as a gate electrode for the graphene, also exhibits substantial modulation of the transmitted THz radiation under applied voltage biases.

**Introduction** Monolayer graphene presents the exciting possibility of an atomically thin material that can be used for a wide range of electronic and optoelectronic applications. Its atomically thin nature, however, presents serious design constraints in these applications. For example, since each layer of graphene can only absorb 2.3% of the incident light, the optical density of these devices will be very small. In addition, graphene's gapless nature has attracted a lot of attention for its use in the far infrared wavelength range, where it has not yet been possible to build optoelectronic devices that are commonly taken for granted in the visible and near infrared wavelength ranges.

In 2012, Rodriguez *et al.* presented the first experimental demonstration of a THz modulator based on the gate tuning of the Fermi level in single layer graphene. Interestingly, they found a modulation depth of 15% but only in a very narrow bandwidth <100GHz centered at 570GHz.[1] Gao *et al.* reported THz modulation up to 50% by exploiting the transmission through ring-shaped metallic apertures placed on top of voltage-gated single layer graphene. Despite the large modulation depth, however, the bandwidth was limited to 0.25 THz, due to the resonance peak at 0.44THz, which suppresses any off-resonance signal, not to mention the complicated fabrication process of the device.[2] In other related work, Maeng *et al.* used terahertz time domain spectroscopy (THz-TDS) to probe the Dirac fermion dynamics in single layer graphene, and found nonlinear and constant behaviors in the sheet conductivity for high and low densities of carriers, respectively. However, the maximum conductivity they measured was only 2.1 mS/square,[3] which is more than one order of magnitude smaller than the trilayer graphene (TLG) sheet conductance presented here (i.e., 57mS/square). Most recently, Li *et al.* reported a large THz modulation depth of 83% using double layer graphene on silicon by simultaneous optical excitation. While this depth of modulation is impressive, the additional light source used to excite the sample introduces difficulties for practical application.[4] One open question that remains unanswered in these previous studies is whether the origin of the modulation is exclusively due to the graphene, or if there is a partial contribution from the underlying gated substrate.

In this letter, we investigate the gate voltage dependence of the optical transmission of randomly stacked trilayer graphene (TLG) by means of THz-time domain spectroscopy (THz-TDS). In addition to the $Si/SiO_2/TLG$ stack, we also measure the THz transmission spectra through the underlying silicon substrate without TLG as a reference, which exhibits a surprisingly large modulation of the THz beam with applied voltage. Unlike previous studies, we normalize each TLG spectrum by a different reference spectrum at each voltage.

**Experimental Details:** The samples used in this study consist of large-scale polycrystalline TLG. First, we grow single layer graphene by chemical vapor deposition (CVD) on copper foil.[5] Then, we transfer this monolayer graphene to an oxidized silicon substrate (i.e., $SiO_2/Si/SiO_2$) using a wet transfer method. The $SiO_2$ thickness on both sides of the silicon is 290nm. In this transfer process, the graphene-on-Cu is coated with poly-methyl methacrylate (PMMA), and the Cu is dissolved in an aqueous solution of copper etchant. The PMMA/graphene stack is then transferred from the solution to the $SiO_2/Si/SiO_2$ substrate. The PMMA is subsequently dissolved by acetone, and the sample is immersed in deionized water to remove any organic residue.[5] This transfer process was repeated to stack three layers of graphene on the silicon substrate. In this configuration, the silicon was used as a back gate, allowing us to electrostatically tune the charge-carrier density in the TLG. Gold electrodes were deposited on the graphene using electron beam metal deposition. A part of the $SiO_2/Si/SiO_2$ substrate was left intentionally uncovered by graphene in order to be used as reference. The size of our TLG sample was approximately 10x10mm$^2$, as shown in Figure 1a. The number of layers was confirmed by their optical contrast, as shown in the inset of Figure 1a, which plots the spatial profile of the reflected light intensities that correspond to different graphene layer thicknesses. Here, the orientation of one layer of graphene with respect to another is random, and we assume that our sample does not possess any Bernal ABA or ABC (rhobohedral) crystallographic stacking. Thus, we treat our trilayer sample as equivalent to single layer graphene with a linear, gapless band structure. Figure 1b shows the Raman spectrum of this sample, which is consistent with monolayer graphene, as expected for randomly stacked TLG.

Our THz-TDS setup is driven by a 5 kHz Ti:Sapphire regenerative amplifier with 2mJ and 60 fs pulse duration at 800 nm. A part of the output is used to generate THz pulses via optical rectification in a 1mm thick ZeTe (110) nonlinear crystal.[6] The emitted THz radiation was focused on the TLG sample with a parabolic mirror at normal incidence. The spot size of the THz beam on the sample was about 0.7mm, measured by the knife-edge method. The transmitted radiation through the sample was collected and refocused by a pair of parabolic mirrors onto another 1mm thick ZnTe (110) crystal, and detected by free-space electro-optic sampling.[6-8] A portion of the laser beam is used as a sampling beam that is scanned via an optical delay stage and is used to sample the temporal electric field profile of the THz transients. All of the measurements were performed at room temperature in a dry environment obtained by enclosing the THz spectrometer in a box purged with nitrogen gas. A lock-in amplifier, phased-locked with a chopper used to modulate the THz beam at 1.5KHz, was used to collect the signal.

The time-domain electric-fields (denoted as $E_{gr}(t)$ and $E_{ref}(t)$) of the THz pulses are transmitted through the sample (TLG deposited on $SiO_2/Si/SiO_2$) and through the reference substrate ($SiO_2/Si/SiO_2$ without TLG) are shown in Fig. 2(a). We windowed the time-domain data to remove the etalon pulses that are well separated from the main pulse. From our time-domain data, we directly observe a large attenuation of the THz transmission, as a result of the presence of charge carriers in the trilayer graphene. The THz field complex transmission coefficient $T^*(\omega)$ is obtained from the ratio between the two Fourier transformed spectra $E_{gr}(\omega)$ and $E_{ref}(\omega)$. We derive the complex transmission coefficient as $T^*(\omega) = |T(\omega)| e^{i\varphi(\omega)}$, where $|T(\omega)|$ and $\varphi(\omega)$ are the amplitude and the phase, respectively. Applying the thin-film approximation,[9-12] we have

$$T^*(\omega) = \frac{E^*_{gr}(\omega)}{E^*_{ref}(\omega)} = \frac{1+n}{1+n+NZ_0\sigma(\omega)} \quad (1)$$

We can extract the frequency-dependent complex sheet conductivity $\sigma(\omega) = \sigma_1(\omega) + i\sigma_2(\omega)$ of the TLG. In Equation 1, $n = 3.42$ is the refractive index of the Si layer because the THz wavelength is sufficiently long compared to the thickness of the $SiO_2$, $N$ is the number of the graphene layers, and $Z = 377\Omega$ is the vacuum impedance. By inverting Eq. 1, we can obtain the real and imaginary parts of the complex conductivity, $\sigma_1$ and $\sigma_2$, from the experimental data as

$$\sigma_1 = \frac{n_{Si}+1}{Z_0}\left(\frac{\cos[\varphi(\omega)]}{|T(\omega)|}-1\right) \quad \text{and} \quad \sigma_2 = -\frac{(n_{Si}+1)}{Z_0}\frac{\sin[\varphi(\omega)]}{|T(\omega)|}. \tag{2}$$

**Results and discussion:** Figure 2(a) shows the time-domain THz electric field transmission with and without the TLG sample. Figure 2(b) shows the Fourier transformed spectrum of the transmitted THz radiation through the substrate and the graphene samples. Here, significant attenuation is observed in the graphene sample spectra due to the presence of the charge carriers. We also measured the transmission of the THz radiation through the reference substrate $T_{ref}^{(V_g)}$ as a function of the applied gate voltage ($V_g$). Here, we observe a strong gate-voltage dependence of the THz transmission through the reference substrate (Fig.2c). This effect has not been reported in previous studies related to THz modulation using gated-graphene devices. In order to remove any residual modulation of the transmitted THz radiation caused by the gated substrate, we normalized our THz transmission spectra measured through graphene under various gate voltages by those obtained from the reference substrate under the same gated voltages. Again, this method is applied for first the time, in contrast with previous studies in which the observed modulation possibly contains modifications of the transmission spectra resulting from the gated substrate itself.

Another interesting observation is the large attenuation of the THz radiation passing through the TLG. Figure 2d shows an 82% change in the peak THz electric field amplitude transmitted through the TLG sample compared to that transmitted through the reference substrate at zero applied voltage. This results from the high density of the charge carriers in the TLG. The carrier densities found by fitting the conductivity data in Fig. 3 with the Drude model are on the order of $10^{13} cm^{-2}$. Here, the observed phase-shift of the time domain THz fields is relatively small, so the imaginary part can be neglected. While the data presented in Figure 2(d) shows a shift of approximately 400fs, most of the samples show negligible shifts.

Figure 3(a) shows the negative differential transmission defined as $[1-T_{gr}^{(V_g)}/T_{ref}^{(V_g)}]$, where $T_{gr}^{(V_g)}$ and $T_{ref}^{(V_g)}$ are the transmission through the graphene plus reference and reference only, respectively. Here, the spectra are almost flat in the THz spectral window allowed by our setup, and the differential transmission is as high as 80% at zero applied gate voltage.[13-16] The corresponding high absorption is due to the large number of carriers and available density of states in graphene. The transmission varies with the applied gate voltages by 15% over a very broad spectral range from 0.5 – 1.6 THz. A similar percentage modulation has been observed from other groups but with a much narrower bandwidth (<0.1THz, Rodriguez *et al.*). A large modulation amplitude has also been reported based on laser-pumped Si devices.[4] The advantage of our method for modulating the THz radiation, however, is the simple fabrication process of the device and that no additional light source is required in order to obtain the modulation. Figure 3(b) shows the real part of the complex-sheet conductivity for a variety of different gate voltages of the TLG in the THz spectral range (from 0.6 to 1.6 THz). The intraband conductivity in graphene is modeled with a semiclassical Drude model with a finite temperature as[17-19]

$$\sigma(\omega) = \frac{2e^2}{\pi\hbar^2} k_B T \cdot \ln\left[2 \times \cos\left(\frac{E_F}{2k_B T}\right)\right] \cdot \frac{i}{\omega + i\tau^{-1}}. \tag{3}$$

For $k_B T \ll E_F$, σ is proportional to $E_F$ and Equation 3 is reduced to the Drude model σ(ω) = $\sigma_{DC}(E_F)/(1+i\omega\tau)$, where $\sigma_{DC}(E_F)$ corresponds to the DC conductivity and $\tau$ is the scattering time. This equation suggests that tuning of the Fermi level reflects corresponding changes in conductivity. The sheet conductivity varies with the applied gate voltage from 1.2 x 10$^{-2}$ to 2.2 x 10$^{-2}$ in units of *($\pi e^2/2h$)*. This large conductivity is expected because of the high density of carriers arising from unintentional doping, but mostly due to the presence of three graphene layers stacked on top of each other. Our results also show a flat response for all the applied voltages in good agreement with those reported for single layer graphene in several studies[3, 20-21] suggesting that the scattering rates of the carriers are high enough that there is not any observable frequency dependence. From the fits of our data, we deduced scattering times in the range of 90 to 200 fs and carrier mobilities of ~3000 cm$^2$V$^{-1}$s$^{-1}$. In general, there is a large mobility variation in monolayer graphene due to its sensitivity to charged impurities. Here, we anticipate a lower degree of variation in the mobilities and increased screening due to the higher density of carriers and the layered nature of our sample, respectively. Thus, the conductivity will be less sensitive to the variations in the mobility of the charge carriers, making the TLG a more suitable material for charge transport applications.

Each transmission spectrum in Figure 3 has been averaged, since the conductivity is not frequency dependent and plotted in Figure 4(b) as a function of the applied gate voltage. By changing the gate voltage, the carriers injected in the graphene layers cause a shift of the Fermi level resulting in an increase or decrease of the THz transmission depending on the position of the Fermi level with respect to the Dirac cone. Depending on the polarity of the gate voltage, holes (*p*-doping) or electrons (*n*-doping) can be injected into the graphene causing the Fermi level to be shifted into the valence or conduction band, as shown in Figures 4(a-d). As the Fermi level shifts to a position that allows a larger number of transition[22] (higher density of states), as shown in Figure 4a, the THz absorption increases and the transmission decreases. When the Fermi level approaches the Dirac cone (Fig. 4b), the available phase-space for intraband transitions is reduced, and the density of states is lowered, resulting in a higher transmission of the THz radiation. A similar process occurs in the conduction band having applied positive gate voltages (i.e., *n*-doping). The conductivity at the peak absorption is 57mS, and the corresponding charge carrier density is estimated to be approximately $1.1 \times 10^{13}$ cm$^{-2}$, as estimated by using a simple capacitor model. For our sample, the capacitance is calculated from the relation $C=\varepsilon_r \varepsilon_0/d=$ 12nF/cm$^{-2}$, where $\varepsilon_r$ =3.9 for SiO$_2$, $\varepsilon_0$ = 8.85×10$^{-12}$ F/m$^2$ is the permittivity of free space, and $d$ = 290 nm is the thickness of the dielectric SiO$_2$.

In order to quantitatively describe the modulation depth (defined as [T(V$_g$)-T(20)]/T(20)), the average transmission data normalized by the lowest absorption, which was obtained at 20V, have been plotted as function of the applied gate voltage. The data in Figure 4(b) are fitted with a Lorentzian function. The peak value near zero gate voltage corresponds to the Dirac point, where THz absorption via intraband transmissions is minimal.

Figure 5 shows calculations of the differential transmission [1-$T_{Gr}$/$T_{ref}$] based on Eq.1 plotted as a function of the number of graphene layers for different conductivity values, consistent with previous values from literature [1,23,24]. These theoretical calculations are plotted together with the experimental values of the differential transmission, taken as the average values of the differential transmission for each spectrum shown in Figure 3a. Here, the experimental data points fall within the range of the theoretical values (solid curves) indicating that our differential transmission results agree well with the theory. Since our sample consists of trilayer graphene, all of the experimental data points in Figure 5 correspond to N=3. The calculations show that beyond 3-layers of graphene (N=3) the differential transmission does not show significant changes, especially at high doping levels.

In conclusion, we fabricated CVD-grown artificially stacked trilayer graphene devices to modulate the transmission of THz radiation. We observed i) a strong absorption >20% along with higher density of carriers ($10^{13}$ cm$^{-2}$) and ii) a modulation depth of 15% over a broad spectral range (0.6-1.6 THz) by varying the applied gate voltage without the use of any additional external photoexcitation. Our approach shows that artificial multilayer graphene preserves all the properties of single layer graphene and provides stronger THz absorption due to the additional graphene layers. Moreover, the underlying silicon substrate also exhibits substantial modulation of the transmitted THz radiation under applied voltage biases, and these effects have been normalized out in the data analysis.

**Author information**

**Corresponding Authors**

E-mail: ioannis.chatzakis.ctr.gr@nrl.naval.mil

Present address: Naval Research Laboratory 4555 Overlook Ave SW Code 6881 Washington, DC 22375 Bldg. 208/231

E-mail: scronin@usc.edu.

**Acknowledgements:**

This work was supported by the Ming Hsieh Institute and a Provost's Postdoctoral Scholar Research Grant at the University of Southern California. This research was supported in part by NSF Award No. CBET-1402906 (I.C), Department of Energy (DOE) Award No. DE-FG02-07ER46376 (Z.L.), and NSF Award No. CBET-1512505 (A. B.).

**Author contribution**

S.B.C. conceived and supervised the experiments; I. C. developed the experimental setup hosted in A.B. laboratory, and performed the measurements; Z. L. provided the samples; I. C. and S.B.C. analyzed the data; I.C. and S.B.C. wrote the paper. All authors discussed the results.

**Figures**

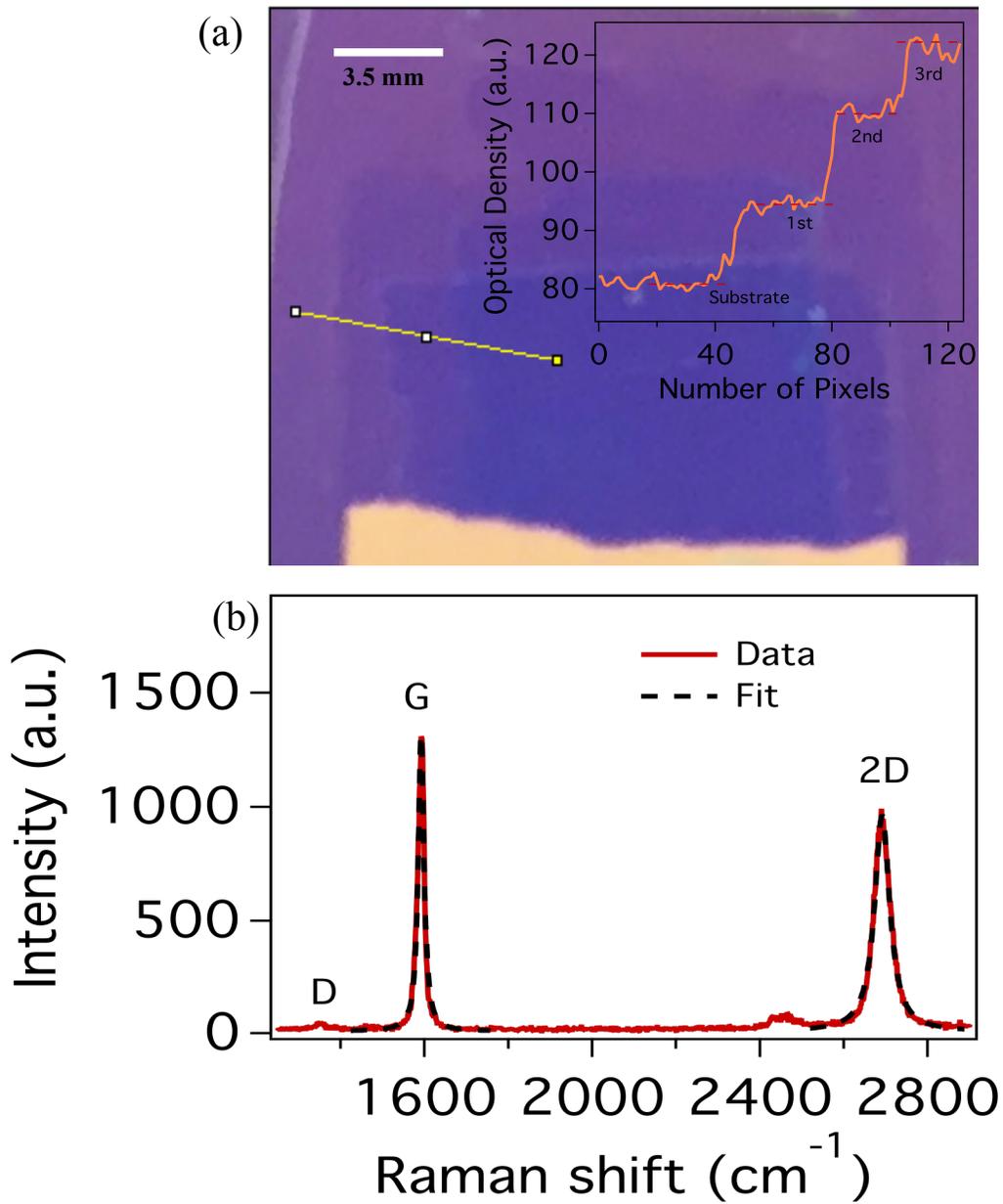

**Fig. 1** (a) Photograph of the TLG on Si/SiO$_2$ device. The inset shows the thickness profile of the TLG on the substrate over the region indicated by the yellow line. (b) Raman spectrum of the TLG shown in (a). The main peaks correspond to the G and 2D modes.

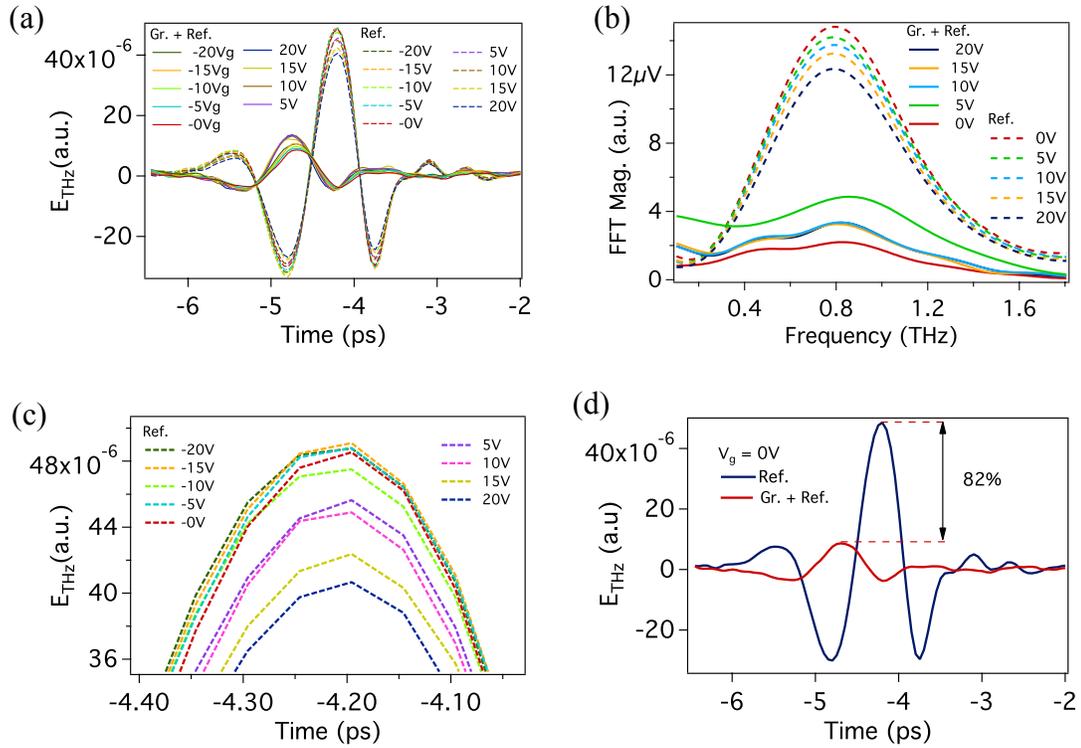

**Fig. 2** (a) Time-domain THz waveforms measured at different applied gate voltages transmitted through the TLG sample and the reference substrate. (b) Fourier-transformed spectra of the corresponding time-domain data plotted in (a). (c) Expanded THz waveforms from (a) transmitted through the gated substrate, clearly shown a large attenuation of the electric field transmission in the substrate. (d) The percentage difference between the amplitudes of the THz waveforms transmitted through the TLG and the substrate. One can observe a small phase shift on the order of few hundreds of femtoseconds.

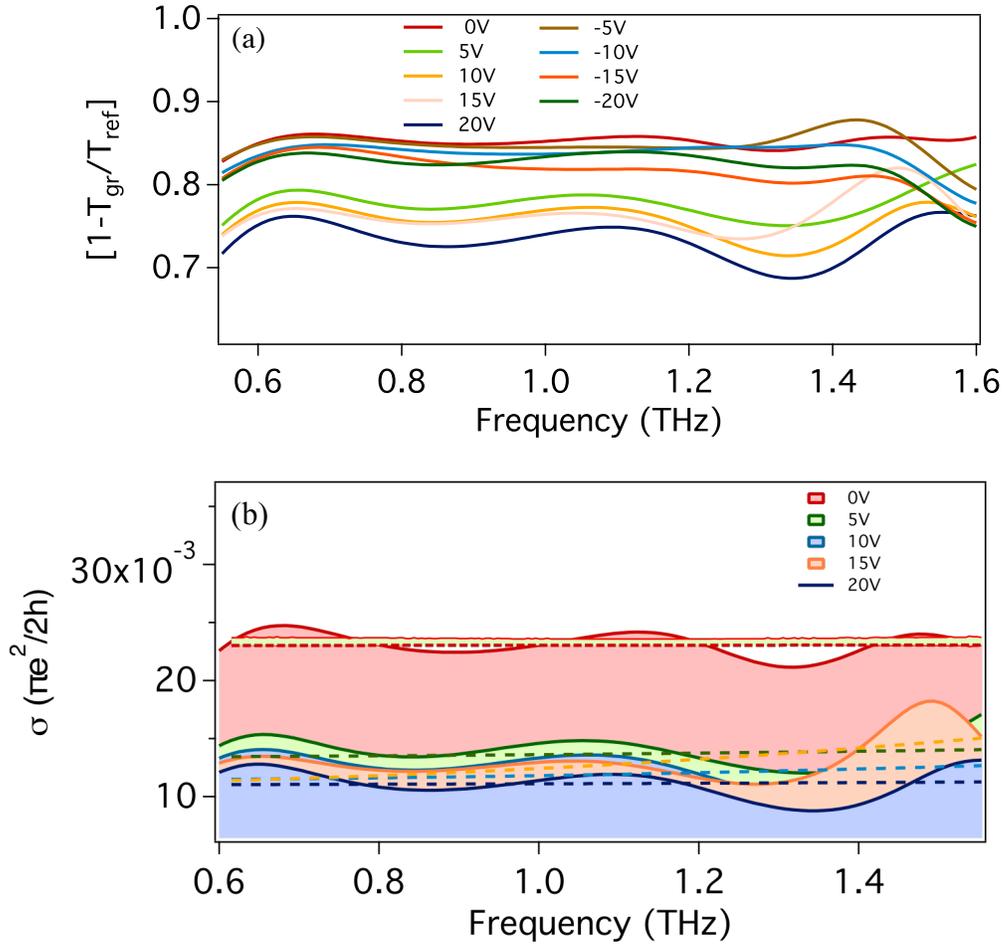

**Fig. 3** (a) Average differential transmission (1-$T_{gr}/T_{ref}$) obtained at different applied gate voltages plotted as a function of frequency. (b) The corresponding intraband conductivity for a few selected spectra extracted using Eq.1 in the text. The dashed lines are the fits to the Drude model, while the thin solid red lines show the confidence interval for the zero Volt conductivity fit.

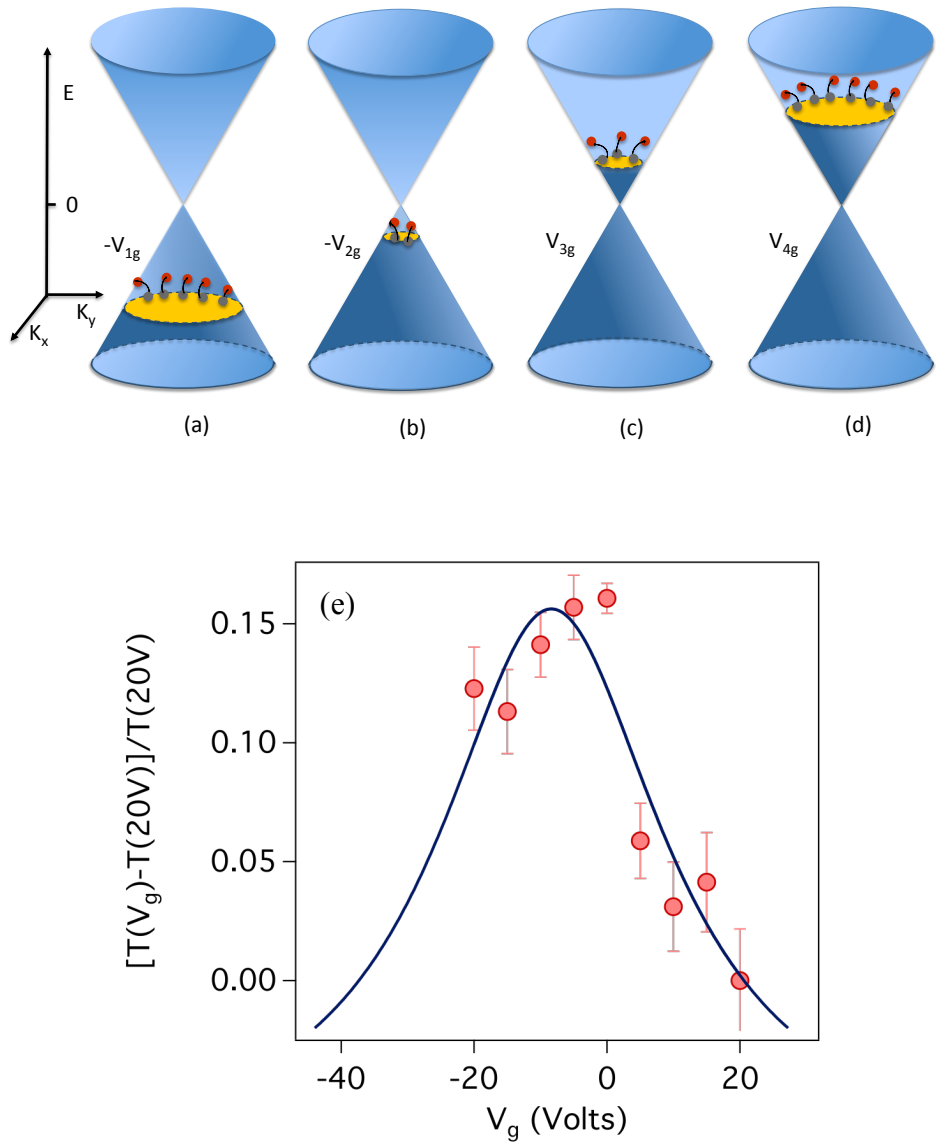

**Fig. 4** (a-d) Energy band diagrams illustrating the shift of the Fermi level (yellow) in the valence and conduction bands depending on the applied gate voltages. The yellow arrows indicate the transitions excited by THz photons. (e) The modulation depth plotted as a function of the applied gate voltage. Here, the data have been normalized by the transmission at gate voltage 20V, where the minimum absorption has been observed. The blue line is a fit to a Lorentzian function.

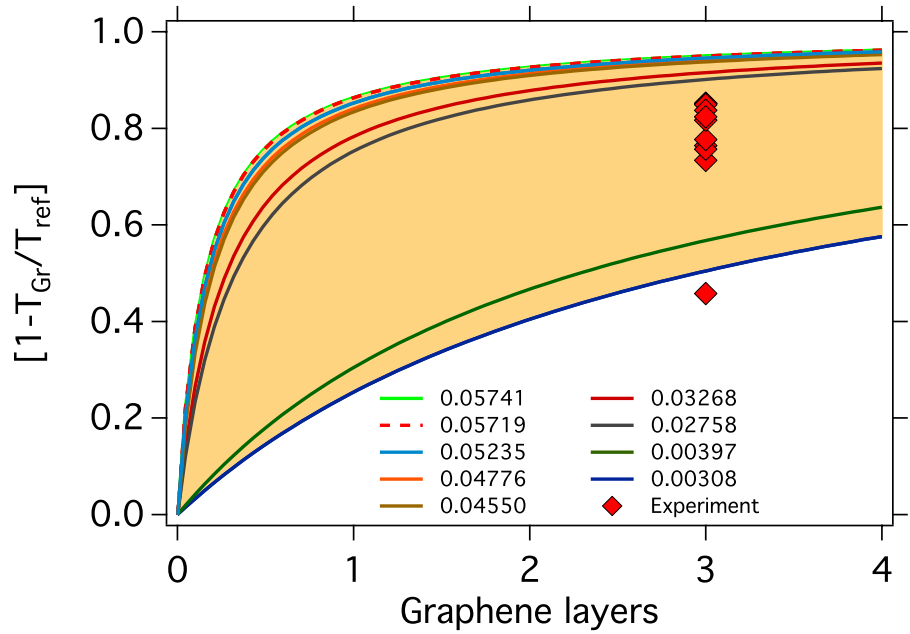

**Figure 5.** The differential transmission [1-$T_{Gr}/T_{ref}$] plotted as a function of the number of graphene layers for different conductivities based on Eq. (1). The conductivity values in the legend are given in units of $\pi e^2/2h$. The red diamonds show the experimental average values of the differential transmission for each spectrum shown in Figure 3a.